\documentclass[osajnl,preprint,showpacs]{revtex4}
\DeclareRobustCommand{\baselinestretch{2.2}}

\usepackage{amsmath}
\usepackage{amssymb}
\usepackage{graphicx}
\usepackage{bm}
\usepackage{mathrsfs}

\begin{document}

\title{Statistical description of short pulses in long optical fibers: Effects of
nonlocality}  

\author{Padma K.\ Shukla and Mattias Marklund}
\affiliation{Department of Physics, Ume{\aa} University, SE--901 87 Ume{\aa},
Sweden} 
\affiliation{Institut f\"ur Theoretische Physik IV, Fakult\"at f\"ur
  Physik und Astronomie,  Ruhr-Universit\"at Bochum, D--44780 Bochum,
  Germany} 

\date{\today}

\begin{abstract}
We present a statistical description of the propagation of short pulses 
in long optical fibers, taking into account the Kerr and nonlocal 
nonlinearities on an equal footing. We use the Wigner approach on 
the modified nonlinear Schr\"odinger equation to obtain a
wave kinetic equation and a nonlinear dispersion relation. The
latter exhibit that the optical pulse decoherence reduces the
growth rate of the modulational instability, and thereby
contribute to the nonlinear stability of the pulses in 
long optical fibers. It is also found that the 
interaction between spectral broadening and nonlocality tends
to extend the instability region. 
\end{abstract}
\ocis{030.1640, 060.2310, 060.4370, 190.7110}

\maketitle

The nonlinear propagation of pulses in optical fibers has attracted
a great deal of interest since the early seventies \cite{Hasegawa,Hasegawa2},
and research into the application and theory of this field is still 
growing \cite{Kivshar}. The evolution of coherent weakly nonlinear 
optical pulse envelopes is given by the cubic  nonlinear Schr\"odinger equation 
(CNLSE) \cite{Hasegawa,Kivshar} involving the Kerr nonlinearity. The CNLSE 
admits bright, dark and gray solitons, which are used for ultrahigh-speed 
optical communications without pulse broadening and spectral dilution \cite{Has}. 
Even so, it has become clear that the effects of noise in fibers 
and amplifiers could alter the pulse properties in significant ways. 
Thus, it is of crucial importance to obtain qualitative and quantitative 
estimates of the effects of different types of incoherent perturbations 
\cite{Kivshar1,Kivshar2} on the optical pulse propagation.  Recently, the Wigner 
transform technique \cite{Wigner,Tito} in nonlinear dispersive media  
has been used to analyze  Landau-like damping \cite{Fedele,Hall-etal}, 
partially coherent higher order dispersive effects
\cite{Helczynski-etal}, the importance of the incoherence spectrum
\cite{Anderson-etal}, and the influence of incoherence on the 
modulational instability \cite{Anderson-etal2,Hall} for cases involving 
the cubic Kerr nonlinearity.

However, there are other important nonlinearities \cite{Hasegawa0,Hasegawa1,Shukla} 
(e.g. saturation and higher order nonlocal nonlinearities) which can compete with the cubic nonlinearity 
in optical fibers. The combined influence of the cubic and nonlocal nonlinearities 
on the modulational instability of a constant amplitude coherent optical pulse has 
been examined by Shukla and Rasmussen \cite{Shukla}. In this Letter, we  present a 
statistical description of partially incoherent pulses in long optical fibers, 
taking into account the Kerr and nonlocal nonlinearities on an equal footing. 
We use the Wigner approach and deduce a wave kinetic equation from which 
a nonlinear dispersion relation (NDR) has been derived. The NDR is then analyzed to 
demonstrate the effect of random noise on the modulational instability of incoherent 
optical pulses. It is found that the optical pulse decoherency can contribute to
the nonlinear stability of pulses in optical fibers. 

Given the electric field $E(z,t)\exp(ik_0z - i\omega_0t)$ of the optical pulses, 
the evolution the pulse envelope $E$ in the slowly varying envelope limit, 
i.e.\  $k_0 \gg (\partial_z - 2k_0^{\prime}\partial_t)$, is governed by \cite{Shukla}
\begin{equation}\label{eq:nlse}
  i(\partial_z + \Gamma)E + \alpha\partial_t^2E + \beta IE + i\gamma\partial_t(IE) = 0 ,
\end{equation}
where
we have introduced the parameters $\alpha = -k_0^{\prime\prime}/2$,
$\beta = n_2k_0/n_0$, $\gamma = 2n_2/c$, and $\Gamma = k_0\chi_0/n_0$. 
Moreover, the prime denotes differentiation with respect to $\omega_0$, the intensity 
parameter is
given by $I = |E|^2$, the refractive index is $n(\omega_0,I) = n_0 + i\chi_0 + n_2 I$, $n_0 = n(\omega_0)$, and 
and $\chi_0 = \chi(\omega_0)$ represents losses in the medium.

In order to take the effects of partial coherence into account,
we define the space-time correlation function for the electric field as
$C(z_+,z_-,t_+,t_-) = E^*(z_+,t_+)E(z_-,t_-)$, where $z_{\pm} = z \pm \zeta/2$ and
$t_{\pm} = t \pm \tau/2$. Then, the Wigner distribution function of the optical pulse 
is given by \cite{Mendonca}
\begin{equation}\label{eq:spacetimewigner}
  F(z,t,k,\omega) = \frac{1}{(2\pi)^2}\int\,d\zeta\,d\tau\,
  e^{i(k\zeta - \omega\tau)}C(z_+,z_-,t_+,t_-) ,
\end{equation}
such that
\begin{equation}\label{eq:intensity}
  I(z,t) = \frac{1}{(2\pi)^2}\int\,dk\,d\omega\, F (z,t,k,\omega) .
\end{equation}

Thus, from Eq.\ (\ref{eq:nlse}) the evolution equation for the Wigner function (\ref{eq:spacetimewigner}) corresponding to the envelope field $E$ becomes
(see also Ref.\ \onlinecite{Besieris-Tappert})
\begin{eqnarray}
2\omega\alpha\partial_t F - \partial_z F + 
    2\beta I \sin\left( \tfrac{1}{2}\stackrel{\leftarrow}{\partial_t}%
    \stackrel{\rightarrow}{\partial_{\omega}} \right) F 
    + \gamma\left\{ -  
      \partial_t\left[ I \cos\left( \tfrac{1}{2}\stackrel{\leftarrow}{\partial_t}%
    \stackrel{\rightarrow}{\partial_{\omega}} \right) F \right] 
      + 2\omega I\sin\left( \tfrac{1}{2}\stackrel{\leftarrow}{\partial_t}%
    \stackrel{\rightarrow}{\partial_{\omega}} \right) F  \right\}
    =  2\Gamma F ,
\label{eq:Wigner2}
\end{eqnarray}
where we have performed the Wigner transformation over the time domain.
Here the arrows denotes direction of operation, and the operator
functions are defined in terms of their respective Taylor expansion. 
The system of equations (\ref{eq:intensity}) and (\ref{eq:Wigner2}) 
determines the evolution of short partially
coherent optical pulses in nonlinear media.

In order to analyse the modulational instability and the effects of the 
terms due to a nonzero $\gamma$, we make the ansatz
$F(z,t,\omega) = F_0(\omega) + F_1(\omega)\exp(iKz - i\Omega t) + \mathrm{c.c.}$, where 
$\mathrm{c.c.}$ denotes the complex conjugate, and $|F_1| \ll F_0$.
Moreover, since we are interested in the short-pulse effects, we will for simplicity 
neglect the loss term $\Gamma$ in Eq.\ (\ref{eq:Wigner2}), in order to obtain 
clearly interpretable results.\footnote{It should be stressed that in certain applications,
  the losses may not be small, and the $\Gamma$ term should under
  these circumstances be kept. As noted by Shukla and Rasmussen \cite{Shukla}
  the effect of the loss term is to damp the pulse according to $\exp(-2\Gamma z)$ 
  as it propagates through the fiber.}
Expanding Eq.\ (\ref{eq:Wigner2}) in terms of this ansatz, and using Eq.\ (\ref{eq:intensity}), 
we obtain
\begin{equation}\label{eq:dispersion}
  1 = \frac{1}{2\alpha\Omega}\int\,d\omega\,\frac{\left[ \beta + \gamma(\omega + \Omega/2)\right]%
    F_0(\omega - \Omega/2) 
    - \left[ \beta + \gamma(\omega - \Omega/2)\right]%
    F_0(\omega + \Omega/2)}{\omega + (K  - \gamma\Omega I_0)/2\alpha\Omega} ,
\end{equation}
where $I_0 = \int\,d\omega\,F_0(\omega)$. Equation (\ref{eq:dispersion})
represents the NDR for a short optical pulse, where the pulse may have spectral 
broadening and partial coherence. 

In the case of a mono-energetic pulse, we have 
$F_0(\omega) = I_0\delta(\omega - \Omega_0)$, where $\Omega_0$ 
corresponds to a frequency shift of the background plane wave solution, 
and the NDR (\ref{eq:dispersion}) gives \cite{Shukla}
\begin{equation}\label{eq:dispersion-mono}
  K = 2(\gamma I_0 - \alpha\Omega_0)\Omega \pm
    \left[ \gamma^2I_0^2\Omega^2 + \alpha^2\Omega^4 
      - 2\alpha I_0(\beta + \gamma\Omega_0)\Omega^2 \right]^{1/2} .
\end{equation}

In practice however, the wave envelope will always suffer perturbations due to
various noise sources, e.g.\ fiber and amplifier noise. A noisy environment may cause the
pulse field to attain a random component in its phase. Thus, 
if the phase $\varphi(x)$ of the electric field varies stochastically, such that the ensemble
average of the phase satisfies \cite{Loudon,Anderson-etal3}
%
$  \langle \exp[-i\varphi(t + \tau/2)]\exp[i\varphi(t - \tau/2)]\rangle = \exp(-\Omega_T|\tau|)$, 
the background Wigner distribution is given by the Lorentzian spectrum
\begin{equation}
  F_0(\omega) = \frac{I_0}{\pi}\frac{\Omega_T}{(\omega - \Omega_0)^2 + \Omega_T^2},
\end{equation}
where $\Omega_T$ corresponds to the width of the spectrum. Then, the NDR (\ref{eq:dispersion}) takes the form
\begin{equation}
  1 = I_0\Omega\frac{2\gamma\left[ K - \gamma I_0\Omega 
    + \alpha\Omega(\Omega_0 - i\Omega_T) \right] - 2\alpha\beta\Omega}%
      {(K - \gamma I_0\Omega + \Omega_0 - i\Omega_T)^2 - \alpha^2\Omega^4} ,
\end{equation}
which has the solution
\begin{equation}\label{eq:sol}
  K = 2\left[ \gamma I_0 - \alpha(\Omega_0 - i\Omega_T) \right]\Omega 
    \pm \left[ \gamma^2I_0^2\Omega^2 + \alpha^2\Omega^4 
      - 2\alpha I_0(\beta + \gamma(\Omega_0 - i\Omega_T))\Omega^2 \right]^{1/2} .
\end{equation}
This solution generalizes the result (\ref{eq:dispersion-mono}) to the case
of a random phase background envelope field. Equation (\ref{eq:sol}) clearly 
shows that the width gives a nontrivial contribution to the NDR. 
We note that when $\gamma = 0$, we may define the growth rate $\kappa$
according to $K = -2\alpha\Omega_0\Omega - i\kappa$, and 
the width $\Omega_T$ then gives rise to a Landau like damping from Eq.\ (\ref{eq:sol}).

When $\gamma$ is non-zero, the growth/damping behavior becomes
considerably more complex, with new instability regions. 
Letting $f =  \gamma^2I_0^2 + \alpha^2\Omega^2 - 2\alpha I_0(\beta + \gamma\Omega_0)$,
and assuming $\Omega_T \ll f/\alpha\gamma I_0$, we obtain the approximate expression
\begin{equation}
  K/\Omega \approx 2\gamma I_0 - 2\alpha\Omega_0 \pm f^{1/2}  + 2i\alpha\Omega_T  
    \pm i\alpha\gamma I_0\Omega_T/f^{1/2}
\end{equation}
from Eq.\ (\ref{eq:sol}). When $\alpha > 0$, and $2\alpha I_0(\beta + \gamma\Omega_0) 
> \gamma^2I_0^2 + \alpha^2\Omega^2$, we have $f < 0$. Denoting the growth rate by 
$\kappa = -\mathrm{Im}(K)$, we obtain 
%
$  \kappa = |f|^{1/2} - 2i\alpha\Omega_T$.
%
Thus, as expected, the coherence spread $\Omega_T$ gives rise to a smaller growth rate for 
the modulational instability of incoherent optical pulses. We note that this instability 
occurs also when $\gamma = 0$. 

\begin{figure}
  \includegraphics[width=.9\columnwidth]{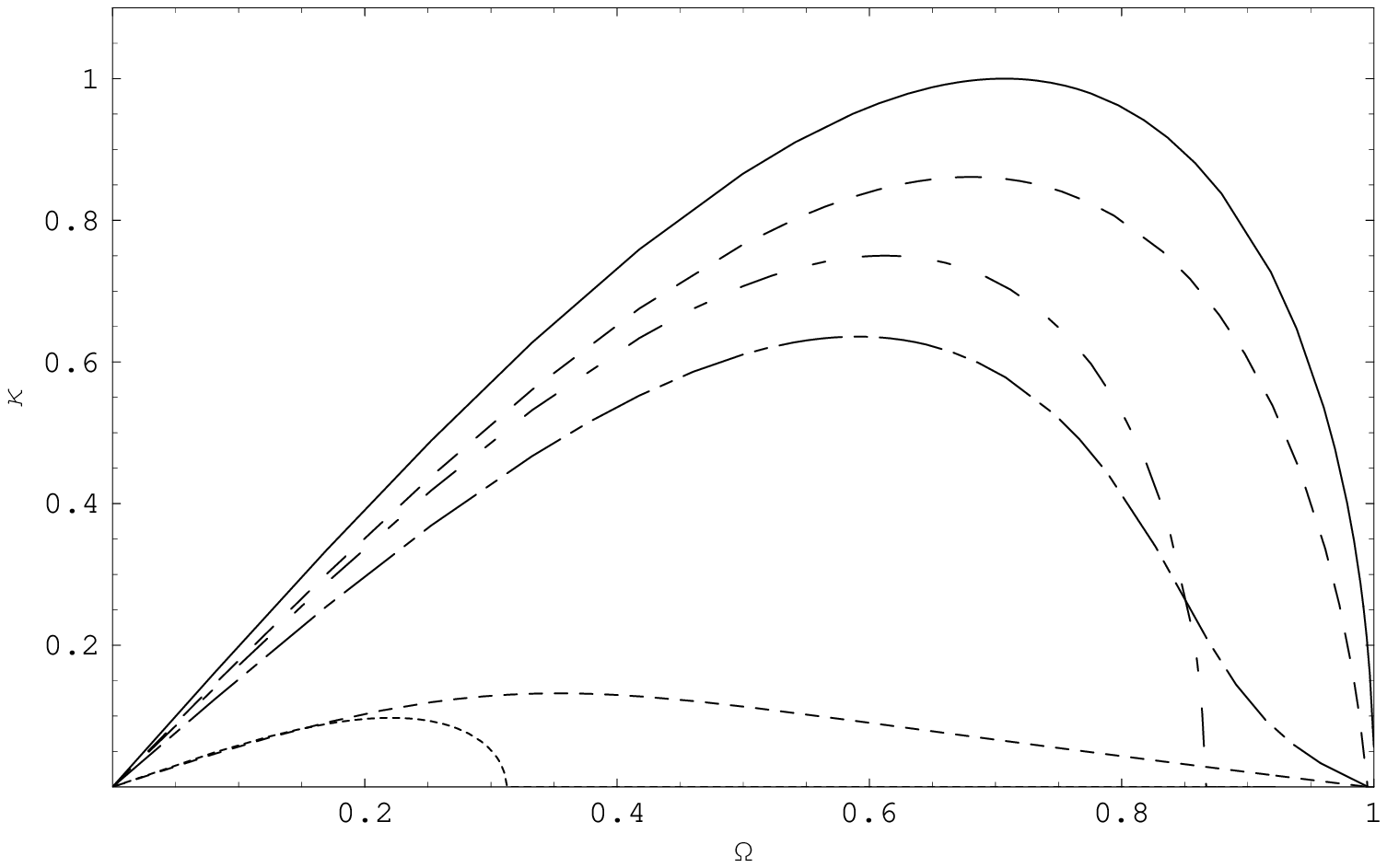}
  \caption{The effects of spectral broadening and non-locality. Normalizing the 
  variables $K$ and $\Omega$, as well as the parameters $\Omega_T$ and $\gamma$, such 
  that $\alpha = \beta = 1$, we have
  plotted the imaginary part $\kappa = -\mathrm{Im}(K)$ as a function of $\Omega$,
  when the frequency shift $\Omega_0$ is put to zero. From the peaks of the 
  curves downwards, we have used $I_0 = 0.5$, and the full curve represents 
  $\Omega_T = \gamma = 0$, and shows the regular
  modulational instability growth rate. The next curve (dashed) gives $\kappa$
  for $\Omega_T=0.1$ and $\gamma = 0$, while the third (dashed-dotted) curve uses $\Omega_T =
  0$ and $\gamma =1$, and the fourth (dashed-dotted) curve has $\Omega_T = 0.1$ and
  $\gamma = 1$. 
  The last two curves (dashed and dotted, respectively), where $\Omega_T = 0.1$, 
  $\gamma = 1.9$, and $\Omega_T = 0$, 
  $\gamma = 1.9$, respectively, clearly shows the character of the 
  combined effect of broadening and non-locality, namely a widening instability region.}
  \label{fig}
\end{figure}

On the other hand, if $\alpha > 0$, 
but $2\alpha I_0(\beta + \gamma\Omega_0) < \gamma^2I_0^2 + \alpha^2\Omega^2$, 
or $\alpha < 0$, so that $f > 0$, a new novel effect is present due to a 
nonzero $\gamma$. We have
%
$  \kappa = \left({\gamma I_0}/{f^{1/2}} - 2\right)\alpha\Omega_T  $. 
Thus, a short pulse in conjunction with a finite statistical spread $\Omega_T$ 
could give rise to a shift in the damping due to the decoherency of the pulse, which 
hence implies a shift also in the growth rate.
This effect can be seen in Fig.\ \ref{fig}, where we have plotted $\kappa$
as given by the full dispersion relation (\ref{eq:sol}) for 
the frequency shift $\Omega_0 = 0$. We have used the rescaling
$I_0 \rightarrow \beta I_0$, $\Omega_T \rightarrow \sqrt{\alpha}\,\Omega_T$,
$\Omega \rightarrow \sqrt{\alpha}\,\Omega$, and 
$\gamma \rightarrow \gamma/(\beta\sqrt{\alpha}) \equiv \sqrt{2}\,n_0/(ck_0|k_0''|^{1/2})$. 
We note that not only is the damping shifted, but the
instability regions is also extended, and quite significantly for higher values of 
$\gamma/(\beta\sqrt{\alpha})$. Since $\gamma/(\beta\sqrt{\alpha}) \propto |D|^{-1/2}$,
where $D$ is the dispersion parameter commonly used in fiber-optics, the value
of the normalized non-locality strength may become large, as $D$ can be 
designed to be very close to zero for certain wavelengths \cite{Agrawal}. Thus, 
the novel coupling between spectral broadening and nonlocality should be possible
to measure using a suitable setup.

To summarize, we have presented an investigation of the modulational instability
of incoherent optical pulses in a nonlinear optical medium that contains the
Kerr and higher order nonlocal nonlinearities on an equal footing. By using the
Wigner transform, we have derived a wave kinetic equation for incoherent pulses
from the generalized nonlinear Schr\"odinger equation. The wave kinetic equation
is further exploited to obtain a nonlinear dispersion relation, which exhibits
new features of the modulational instability. We find that the decoherence of 
the optical pulses reduce the modulational instability growth rate due to a
spatial damping caused by the broad optical pulse spectrum. However,
the combined effect of a random phase and a non-local nonlinearity 
is to extend the instability region as compared to the case of a monochromatic
spectrum. Thus, the present 
result thus contribute to the nonlinear stability of incoherent optical pulses
in long optical fibers.


\newpage

\end{document}